\documentclass[conference]{IEEEtran}
\IEEEoverridecommandlockouts
\usepackage{cite}
\usepackage{amsmath,amssymb,amsfonts}
\usepackage[version=4]{mhchem}
\usepackage{algorithmic}
\usepackage{graphicx}
\usepackage{textcomp}
\usepackage{xcolor}
\usepackage{booktabs}
\usepackage{multirow}
\usepackage{floatrow}
\usepackage[caption=false, font=normalsize, labelfont=sf, textfont=sf]{subfig}
\def\BibteX{{\rm B\kern-.05em{\sc i\kern-.025em b}\kern-.08em
    T\kern-.1667em\lower.7ex\hbox{E}\kern-.125emX}}
\begin{document}

\title{Efficient Simulation of Gate Leakage Currents in Ultra-scaled Transistors from First Principles\\
\thanks{This work was supported by the Swiss National Science Foundation (SNSF) under grant 205602 (NCCR MARVEL) and grant 209358 (QuaTrEx), and by the Platform for Advanced Scientific Computing in Switzerland (BoostQT). We acknowledge support from the Swiss National Supercomputing Center (CSCS) under projects c33, lp16, and lp82.}
}

\author{\IEEEauthorblockN{Mauro Dossena, Nicolas Vetsch, Alexander Maeder, Anders Winka, Vincent Maillou,\\ Denghui Lu, Alexandros Nikolaos Ziogas, Jiang Cao and Mathieu Luisier} \IEEEauthorblockA{Integrated Systems Laboratory (IIS), ETH Zurich, Zurich, 8092, Switzerland\\ e-mail: mdossena@iis.ee.ethz.ch}}

\maketitle

\begin{abstract}
To ensure high electrostatic integrity in ultra-scaled field-effect transistors (FETs), thin gate dielectric layers are typically needed, potentially giving rise to (strong) leakage currents. Being able to model this effect is therefore crucial. Here, we show how gate leakage currents can be computed at the \textit{ab initio} level in WS$_2$-hBN FETs by extending the well-established quantum transmitting boundary method (QTBM). Two main challenges are tackled: (i) the calculation of the open boundary conditions (OBCs) for the gate contacts and (ii) the solution of the resulting sparse linear system of equations \( \mathbf{A} \mathbf{x} = \mathbf{b}\) where the matrix \(\mathbf{A}\) possesses an unstructured sparsity pattern. We demonstrate that the proposed method outperforms the state of the art and can be readily applied to simulate transistors with gate lengths exceeding 25 nm.
\end{abstract}

\section{Introduction}
\begin{figure*}[htbp]
\centering
    \sidesubfloat[]{
        \includegraphics[width=0.6\textwidth]%
        {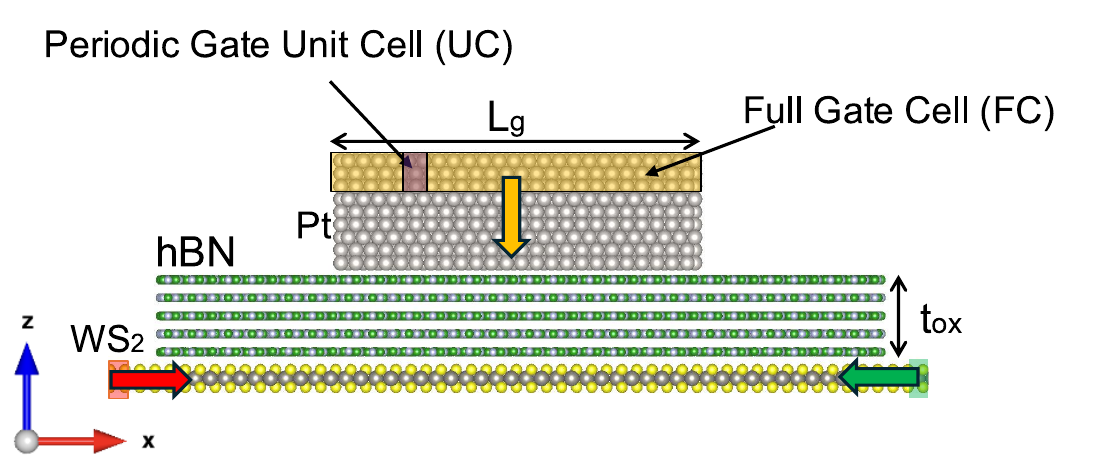}%
    }%
   \sidesubfloat[]{
        \includegraphics[width=0.23\textwidth]{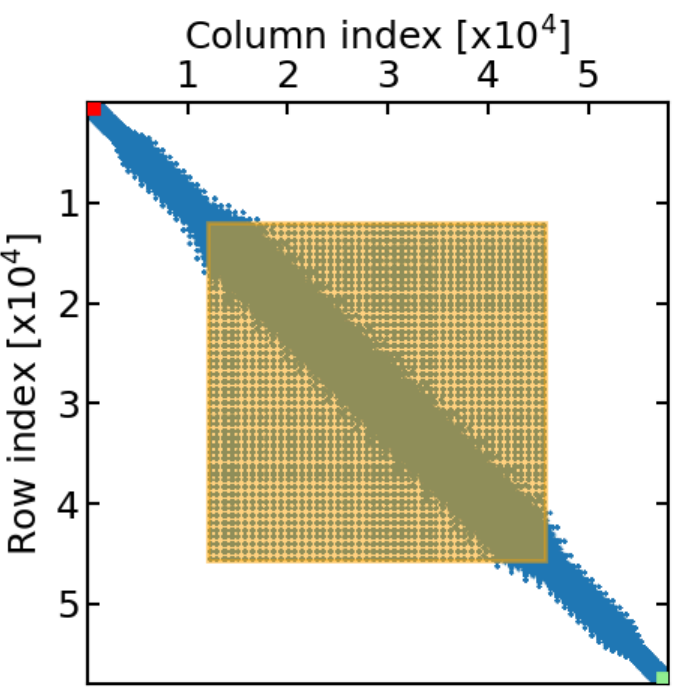}%
    }%
	\caption{(a) Overview of the \ce{WS2}-\ce{hBN}-\ce{Pt} transistor structure considered in this work. It comprises a semiconducting \ce{WS2} channel, a 1.8 nm thick \ce{hBN} insulator layer, and a Pt gate of varying length. Periodic boundary conditions are assumed along the $y$ direction, and transport occurs along the $x$ and $z$ axes. The source, drain, and gate contact regions are highlighted in red, green, and yellow, respectively, while the arrows represent electron injection. The gate contact can either be treated as a single entity (Full Gate Cell, FC) or decomposed into smaller, periodically repeated unit cells depicted in purple (Periodic Gate Unit Cell, UC). (b) Sparsity pattern of the $\left[E\mathbf{S} - \mathbf{H} - \mathbf{\Sigma^B}\right]$ matrix corresponding to the device geometry in (a) (L$_g$ $\approx$ 6 nm) and entering our QTBM solver. Specifically, the entries of $\mathbf{\Sigma^B}$ are marked in red, green, and yellow for the source, drain, and gate contacts.}
	\label{fig1}
\end{figure*}

As a consequence of the continuous miniaturization of the transistor dimensions and of their equivalent oxide thickness (EOT), gate leakage currents could become a major limiting factor in future technology nodes. This issue might specifically affect logic switches with channels made of two-dimensional transition metal dichalcogenides (TMDCs), e.g., \ce{WS2}, for which no intrinsic oxides exist. To shed light on the importance of gate leakage in ultra-scaled devices, an \textit{ab initio} quantum transport (QT) approach is crucial. Although such QT solvers have been used to investigate semiconductor-dielectric interfaces \cite{b1,b2}, to the best of our knowledge, they have not yet been used to study gate leakage. While conventional \textit{ab initio} QT tools typically support carrier injections at two contacts (source and drain) only, accounting for gate leakage currents requires introducing a third injection path at the gate location, which is placed on top of the oxide and extends across the entire channel region. In standard QT approaches, the presence of such a large spatial extent significantly degrades the computational performance because of (i) the calculation of the required open boundary conditions (OBCs) for the gate contact and (ii) the change in sparsity pattern of the resulting linear system of equations. In this work, we address these shortcomings by extending the quantum transmitting boundary method (QTBM) \cite{b3}. The developed approach is applied to the computation of the source-to-gate and drain-to-gate transmission functions of a monolayer WS$_2$ field-effect transistor (FET) with a 1.8 nm thick hBN dielectric layer and a Pt gate metal contact as a representative example (Fig.~\ref{fig1}(a)). The Hamiltonian and overlap matrices of the three different material regions defining this geometry, namely, the contacts consisting only of \ce{WS2}, the buffer area including both \ce{WS2} and hBN, and the central area consisting of \ce{WS2}, hBN, and \ce{Pt}, are determined separately at the density functional theory (DFT) level with CP2K \cite{b4}, whose basis relies on a linear combination of atomic orbitals (LCAO). Subsequently, the three regions are combined to form the complete device Hamiltonian, $\mathbf{H}$, and the corresponding orbital overlap matrix, $\mathbf{S}$, following the procedure outlined in Ref.~\cite{b1}. These device matrices are passed to our open-source QT solver, \textit{quatrex} \cite{b5}, which implements the QTBM to compute the transport properties of multi-terminal nanoelectronic devices.

\section{Methods}

\subsection{Quantum Transmitting Boundary Method}

The QTBM is a computationally efficient framework to evaluate phase-coherent electron transport through nanostructures. It has been widely used, in combination with the effective mass approximation, tight-binding methods, and \textit{ab initio} models. It allows for the incorporation of the effects of semi-infinite leads into the Schr\"{o}dinger equation, giving rise to an equation of the form \cite{b6}
\begin{equation}
    \left[E\mathbf{S} - \mathbf{H} - \mathbf{\Sigma^B}(E)\right]\mathbf{\Phi}(E) = \mathbf{Inj}(E).
    \label{eq:QTBM}
\end{equation}
Here, $E$ is the electron energy at which the system is solved, $\mathbf{\Sigma^B}$ is the open boundary self-energy, capturing the interaction of the system with all connected semi-infinite leads, and $\mathbf{Inj}$ is the injection matrix describing the electron modes entering the device from the semi-infinite leads. The unknown quantity, $\mathbf{\Phi}$, represents the electron wavefunctions generated inside the device.
In Eq.~(\ref{eq:QTBM}), $\mathbf{\Sigma^B}$ is a sparse matrix whose non-zero entries correspond to orbitals directly coupling the central device region to its contact extensions. It can be partitioned as $\mathbf{\Sigma^B} = \mathbf{\Sigma^B_S} +\mathbf{\Sigma^B_D} + \mathbf{\Sigma^B_G}$, where each individual term refers to a contribution from a different contact, i.e., the source, drain, and gate, respectively. The same decomposition applies to the injection matrix $\mathbf{Inj} = \mathbf{Inj_S} + \mathbf{Inj_D} + \mathbf{Inj_G}$. The computation of all terms in $\mathbf{\Sigma^B}$ and $\mathbf{Inj}$ involves the solution of non-linear eigenvalue problems (NEVPs) \cite{b6}, whose time complexity scales with $\mathcal{O}\left(N^{3}\right)$, $N$ being the number of orbitals directly connected to the contacts. A first problem becomes apparent here: Computing $\mathbf{\Sigma^B_G}$ becomes increasingly difficult for long gate electrodes due to the cubic scaling of the NEVP.
The sparsity pattern of the system matrix $\left[E\mathbf{S} - \mathbf{H} - \mathbf{\Sigma^B}\right]$ for a device with L$_g$ $\approx$ 6 nm is shown in Fig.~\ref{fig1}(b), where the contribution of each contact is highlighted with a different color. This illustrates the second challenge: The $\mathbf{\Sigma^B_G}$ term leads to an unstructured sparsity pattern, which drastically increases the fill-in occurring during a direct solution of Eq.~(\ref{eq:QTBM}) via LU factorization. In addition, standard block Gaussian elimination techniques, such as the recursive Green's function algorithm (RGF), do not perform well here due to the large size of the central block induced by the gate contact.

\subsection{Bloch Expansion of Open Boundary Conditions}
To overcome the first limitation, we divide the full cell (FC) gate contact (yellow-shaded region in Fig.~\ref{fig1}) into smaller, repeatable unit cells (UC) that correspond to the purple-shaded region in Fig.~\ref{fig1}. In the case of a periodic contact along two directions, e.g., $x$ and $y$, the UC can be repeated $\alpha$ times along $x$ and $\beta$ times along $y$. Following the Bloch expansion procedure of Refs.~\cite{b7, b8}, we introduce a set of $N_k = \alpha \times \beta$ auxiliary wave vectors $(k_a,k_b)$ sampled over a uniform Monkhorst-Pack grid and compute a UC boundary self-energy $\mathbf{\boldsymbol\sigma^B_G}(k_a,k_b)$ and UC injection matrix $\mathbf{inj_G}(k_a,k_b)$ for each $(k_a,k_b)$ combination. The FC $\mathbf{\Sigma^B_G}$ and $\mathbf{Inj_G}$ are reconstructed as
\begin{gather}
    \mathbf{\Sigma^B_G} = \frac{1}{N_k} \sum_{k_a,k_b}
    \begin{bmatrix}
        1 & e^{-ik_a} & \cdots & e^{-i(\alpha-1)k_a} \\
        e^{ik_a} & 1 & \cdots & e^{-i(\alpha-2)k_a} \\
        \vdots & \vdots & \ddots & \vdots \\
        e^{i(\alpha-1)k_a} & e^{i(\alpha-2)k_a} & \cdots & 1 
    \end{bmatrix} \nonumber \\ \otimes 
    \begin{bmatrix}
        1 & e^{-ik_b} & \cdots & e^{-i(\beta-1)k_b} \\
        e^{ik_b} & 1 & \cdots & e^{-i(\beta-2)k_b} \\
        \vdots & \vdots & \ddots & \vdots \\
        e^{i(\beta-1)k_b} & e^{i(\beta-2)k_b} & \cdots & 1  
    \end{bmatrix} \otimes
    \mathbf{\boldsymbol\sigma^B_G}(k_a,k_b)
    \end{gather}
    \begin{gather}
    \mathbf{Inj_G} = \frac{1}{\sqrt{N_k}}
    \begin{bmatrix}
    \boldsymbol{\gamma}(k_0,k_0) | \boldsymbol{\gamma}(k_0,k_1) | \cdots
\end{bmatrix} \\
\boldsymbol{\gamma}(k_a,k_b) = \begin{bmatrix}
        1 \\
        e^{ik_a}\\
        \vdots\\
        e^{i(\alpha-1) k_a}
    \end{bmatrix}\otimes
    \begin{bmatrix}
        1 \\
        e^{ik_b}\\
        \vdots\\
        e^{i(\beta-1) k_b}
    \end{bmatrix}\otimes \mathbf{inj_G}(k_a,k_b),
    \label{eq:upscale}
\end{gather}
where $\otimes$ denotes the Kronecker product.

Solving the NEVP on a periodic UC grid reduces the computational complexity of the OBC calculation from $\mathcal{O}\left(N^{3}\right)$ down to $\mathcal{O}\left( N_k (N/N_k)^3 \right)$. However, it should be noted that this approach is exact only in the case of a truly periodic contact cell extending infinitely in all transverse directions. In our device configuration, since the gate has finite dimensions along the $x$ direction, this method must be considered an approximation and requires careful validation (see below).

\subsection{Efficient Solution of the System of Equations}
\begin{figure}[t]
\centering
        \includegraphics[width=0.9\textwidth]%
        {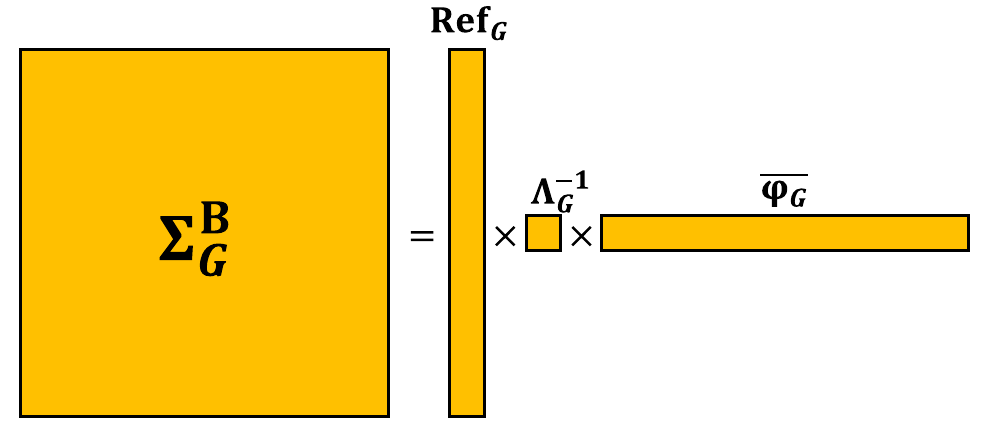}%
	\caption{Overview of the low-rank approximation employed to decompose the gate boundary self-energy $\mathbf{\Sigma^B_G}$ into the product of three matrices, modifying the QTBM numerical problem.}
	\label{fig2}
\end{figure}

The second issue, i.e., the unstructured sparsity pattern of the system matrix, is tackled with a special low-rank treatment of $\mathbf{\Sigma^B_G}$. The $N_G\times N_G$ open boundary self-energy matrix $\mathbf{\Sigma^B_G}$ lends itself to a low-rank decomposition \cite{b6}
\begin{equation}
    \mathbf{\Sigma^B_G} \approx \mathbf{Ref_G} \, \mathbf{\Lambda^{-1}_G} \, \mathbf{\bar{\boldsymbol{\varphi}}_G},
\end{equation}
where $\mathbf{Ref_G}$ is an $N_G \times n_G$ matrix representing the decaying reflected modes at the contact, $\mathbf{\Lambda^{-1}_G}$ is an $n_G \times n_G$ diagonal matrix containing the phase factor of each mode, and $\mathbf{\bar{\boldsymbol{\varphi}}_G}$ is an $n_G\times N_G$ matrix representing the dual (inverse) of every decaying mode. Notably, $n_G \ll N_G$ (Fig.~\ref{fig2}).
Analogous decompositions apply to $\mathbf{\Sigma^B_S}$ and $\mathbf{\Sigma^B_D}$, and can thus be extended to the entire $\mathbf{\Sigma^B}$ matrix. By combining the low-rank approximation with the Woodbury identity \cite{b9}, a modified QTBM equation can be derived
\begin{equation}
    (E\mathbf{S} - \mathbf{H})\begin{bmatrix}
        \mathbf{\Phi'},\mathbf{\Phi''}
    \end{bmatrix} = \begin{bmatrix}
        \mathbf{Inj}, \mathbf{Ref}
    \end{bmatrix},
    \label{eq:QTBM_mod}
\end{equation}
where $\mathbf{\Sigma^B}$ no longer explicitly enters the system matrix, but is replaced by additional columns on the right-hand side of the equation. The solution of the original QTBM problem can be recovered via matrix-vector multiplications:
\begin{equation}
    \mathbf{\Phi} = \mathbf{\Phi'} + \mathbf{\Phi''}\left(\mathbf{\Lambda}-\mathbf{\bar{\boldsymbol{\varphi}}}\mathbf{\Phi''}\right)^{-1}\mathbf{\bar{\boldsymbol{\varphi}}}\mathbf{\Phi'}.
\end{equation}

The system of equations in the modified formulation of Eq.~(\ref{eq:QTBM_mod}) now exhibits a structured sparsity pattern that is more amenable to LU factorization and can potentially be solved via block Gaussian elimination algorithms. Another favorable property of the modified system is that it is Hermitian, requiring only roughly half of the memory and operations compared to the solution of the non-Hermitian system in Eq.~(\ref{eq:QTBM}).
As with the full-rank boundary treatment, the low-rank approximation of $\mathbf{\Sigma^B}$ allows for a Bloch expansion. In this case, the three matrices $\mathbf{Ref}$, $\mathbf{\Lambda^{-1}}$, and $\mathbf{\bar{\boldsymbol{\varphi}}}$ can be upscaled from their UC counterparts, following an approach similar to the upscaling of Eq.~(\ref{eq:upscale}). Combining the two approximations enables the simulation of devices with L$_g$ exceeding 25 nm.

\section{Results}

\begin{figure}[t]
\centering
    \sidesubfloat[]{
        \includegraphics[width=0.85\textwidth]
        {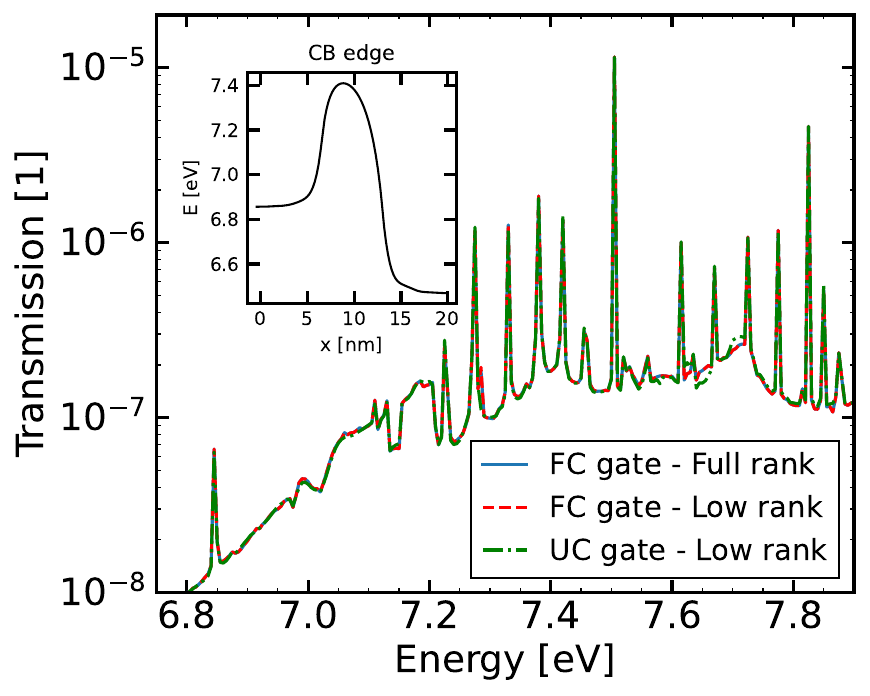}%
    }%
    \\ \vspace{2mm}
   \sidesubfloat[]{
        \includegraphics[width=0.85\textwidth]{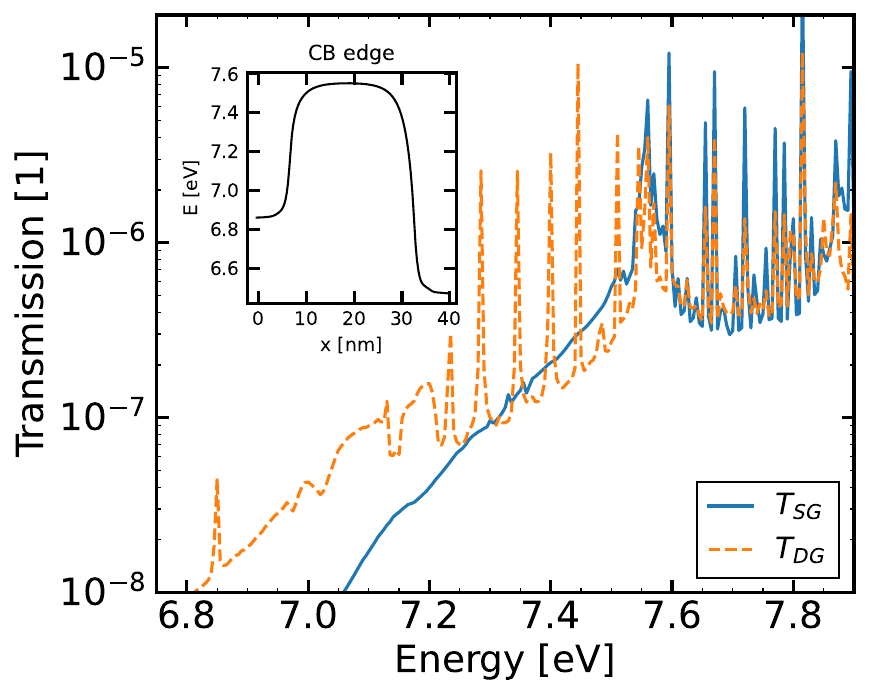}%
    }%
	\caption{(a) Comparison of the drain-to-gate transmission function in the \ce{WS2}-\ce{hBN}-\ce{Pt} transistor from Fig.~\ref{fig1}(a) with a gate length L$_g$ = 6.4 nm, as calculated with the original QTBM approach (blue line), with the low-rank approximation (dashed red line), and with the additional division of the gate electrode into small, repeatable unit cells (dashed-dotted green line). The conduction band (CB) edge of the device is shown in the inset. (b) Source-to-gate (blue) and drain-to-gate (orange) transmission functions for the same \ce{WS2}-\ce{hBN}-\ce{Pt} structure as in (a), but with L$_g$ = 25.6 nm.}
	\label{fig3}
\end{figure}

\begin{figure}[t]
\centering
    \sidesubfloat[]{
        \includegraphics[width=0.85\textwidth]
        {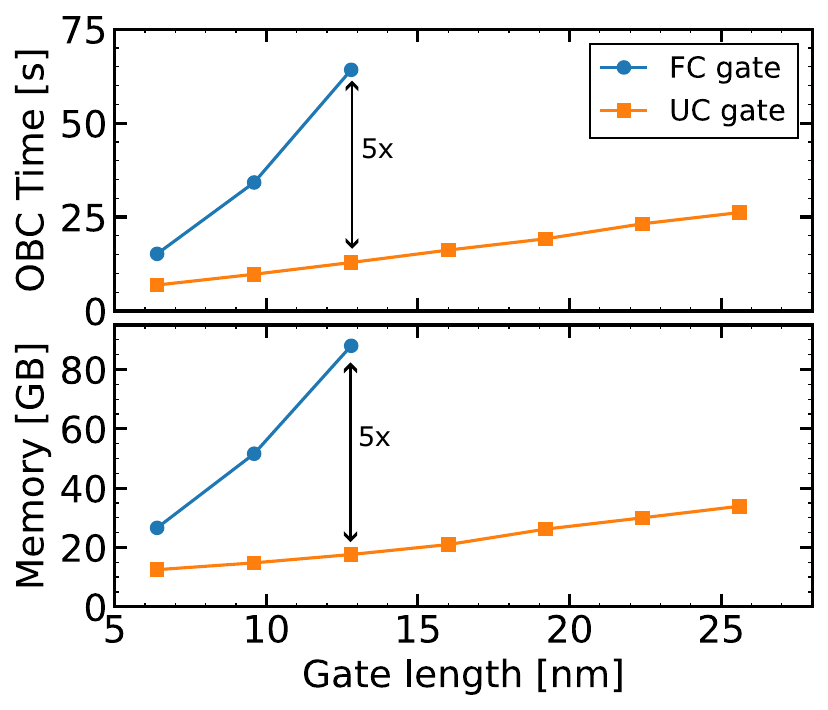}%
    } \\ \vspace{2mm}
   \sidesubfloat[]{
        \includegraphics[width=0.85\textwidth]{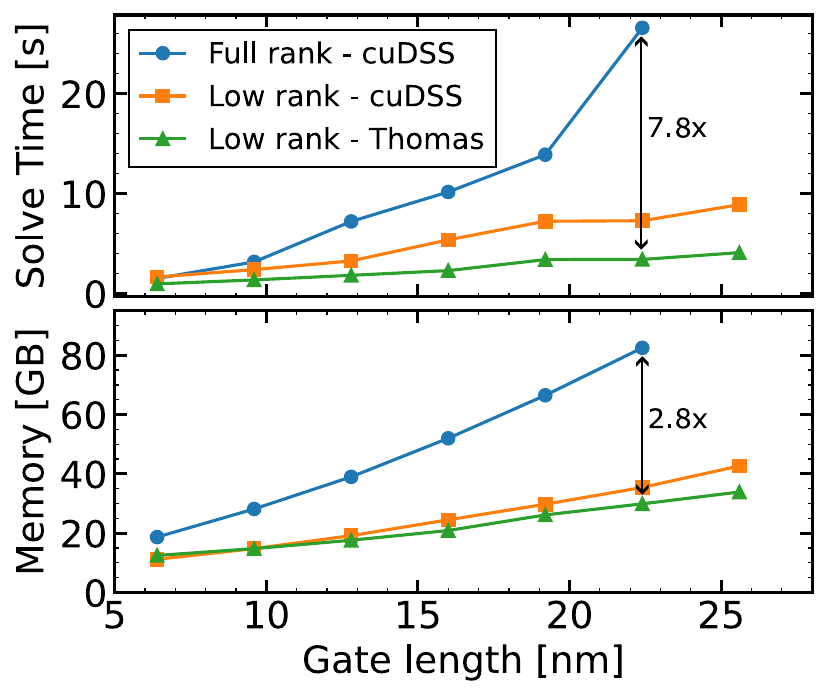}%
    }%
	\caption{(a) Runtime (top) and memory footprint (bottom) on an NVIDIA GH200 GPU during the calculation of the gate boundary self-energy $\mathbf{\Sigma^{B}_G}$ as a function of the gate length for the \ce{WS2}-\ce{hBN}-\ce{Pt} transistor from Fig.~\ref{fig1}(a). The blue lines with circles refer to the case where the entire gate contact is treated as a single block, while the orange lines with squares correspond to the unit cell Bloch expansion method. (b) Same as (a), but for the solution of the QTBM problem. Three solvers are compared: NVIDIA cuDSS using the full-rank $\mathbf{\Sigma^{B}}$ (blue lines with circles), cuDSS when employing the low-rank approximation (orange lines with squares), and a custom implementation of the block Thomas algorithm with low-rank approximation (green lines with triangles).}
	\label{fig4}
\end{figure}

\subsection{Validation of Approximations}
The two approximations introduced in the previous section are examined here. By selecting a device with L$_g$ = 6.4 nm, we can compute a baseline solution without approximations where the NEVP of the full gate cell is solved and the full-rank $\mathbf{\Sigma^B_G}$ in Eq.~(\ref{eq:QTBM}) is used. This baseline is then compared to a solution employing only the low-rank technique and another one exploiting both the Bloch expansion and low-rank OBC approximation. To make the simulation conditions as realistic as possible, an external electrostatic potential bending the channel's conduction band is added to the Hamiltonian, $\mathbf{H}$. It is shown in the insets of Fig.~\ref{fig3}. The resulting drain-to-gate transmission functions for the three aforementioned cases are displayed in Fig.~\ref{fig3}(a). All curves agree excellently, demonstrating the validity of the low-rank and Bloch expansion approximations, even at L$_g$ = 6.4 nm.

Importantly, this method enables the simulation of gate leakage in devices with realistically-sized gate contacts. To illustrate our model's capabilities, the source-to-gate and drain-to-gate transmission functions are computed for a structure with L$_g$ = 25.6  nm, as depicted in Fig.~\ref{fig3}(b). As can be expected from the steeper potential drop on the drain side of the FET, the drain-to-gate transmission function is larger than its source-to-gate counterpart. Both gate leakage transmission functions exhibit distinct peaks around 7.55 eV, which corresponds to conduction band states in the \ce{WS2} channel. Electrons injected into the channel can tunnel into the gate along its entire extent, increasing the overall leakage probability at this specific energy.

\subsection{Numerical Performance}

\begin{table}[t]
\caption{Summary of Numerical Performance Results}
\label{tab:sispad_summary}
\centering
\resizebox{\columnwidth}{!}{%
\setlength{\tabcolsep}{3pt} 
\begin{tabular}{llccccccc}
\toprule
\multicolumn{2}{l}{\textbf{Gate Length (L$_g$ [nm])}} & \textbf{6.4} & \textbf{9.6} & \textbf{12.8} & \textbf{16.0} & \textbf{19.2} & \textbf{22.4} & \textbf{25.6} \\
\midrule
\multicolumn{9}{l}{\textbf{Time (s)}} \\
\midrule
\multirow{3}{*}{OBCs} 
    & Full Cell Gate & 15.2 & 34.2 & 64.2 & -- & -- & -- & -- \\
    & Unit Cell Gate & 6.9 & 9.8 & 12.9 & 16.2 & 19.2 & 23.2 & 26.2 \\
    & Speedup & 2.2$\times$ & 3.5$\times$ & 5.0$\times$ & -- & -- & -- & -- \\
\midrule
\multirow{3}{*}{Solve} 
    & Full Rank - cuDSS & 1.5 & 3.2 & 7.2 & 10.2 & 13.9 & 26.6 & -- \\
    & Low Rank - Thomas & 1.0 & 1.4 & 1.8 &2.3 & 3.4 & 3.4 & 4.1 \\
    & Speedup & 1.5$\times$ & 2.3$\times$ & 4.0$\times$ & 4.4$\times$ & 4.1$\times$ & 7.8$\times$ & -- \\
\midrule
\multicolumn{9}{l}{\textbf{GPU Memory (GB)}\textsuperscript{$\dagger$}} \\
\midrule
\multirow{3}{*}{OBCs} 
    & Full Cell Gate & 26.6 & 51.5 & 88.0 & -- & -- & -- & -- \\
    & Unit Cell Gate & 12.5 & 14.8 & 17.6 & 21.0 & 26.2 & 30.0 & 33.9 \\
    & Memory Saving & 2.1$\times$ & 3.5$\times$ & 5.0$\times$ & -- & -- & -- & -- \\
\midrule
\multirow{3}{*}{Solve} 
    & Full Rank - cuDSS & 18.7 & 28.2 & 39.0 & 52.0 & 66.5 & 82.5 & -- \\
    & Low Rank - Thomas & 12.5 & 14.8 & 17.6 & 21.0 & 26.2 & 30.0 & 33.9 \\
    & Memory Saving & 1.5$\times$ & 1.9$\times$ & 2.2$\times$ & 2.5$\times$ & 2.5$\times$ & 2.8$\times$ & -- \\
\bottomrule
\end{tabular}%
}
\par\vspace{2pt}
{\footnotesize $^{\dagger}$\,Memory values are the peak of the full simulation
(OBCs + solve).}
\end{table}
To complete our analysis, in Fig.~\ref{fig4} we provide performance measurements to illustrate the computational benefits of our approach, in terms of runtime and memory. All benchmarks were run on a single NVIDIA GH200 chip with 96 GB of GPU memory. Figure~\ref{fig4}(a) demonstrates how the Bloch expansion method decreases the simulation time and memory footprint, assuming the low-rank approximation is employed to solve the QTBM system. In particular, explicitly treating the full gate contacts limits simulations to L$_g$ $\leq$ 12.8 nm. Conversely, applying the Bloch expansion approximation, we find more favorable time and memory scaling (up to 5$\times$ savings for L$_g$ = 12.8 nm), enabling OBCs computation in structures with L$_g$ = 25.6 nm. In Fig.~\ref{fig4}(b), the performance impact of the low-rank approximation is reported under the assumption that the Bloch expansion approximation is used when solving the OBCs. We performed simulations with and without the low-rank approximation using NVIDIA's cuDSS sparse direct solver. In addition, with the low-rank approximation, we can employ a custom implementation of the block Thomas algorithm \cite{b9}, which is a block Gaussian elimination algorithm, similar to RGF. Due to the large size of the central matrix block, the solution is not possible in the full-rank QTBM formulation. The low-rank approximation provides further computational benefits, namely runtime and memory savings of 7.8$\times$ and 2.8$\times$, respectively, at L$_g$ = 22.4 nm. Table~\ref{tab:sispad_summary} summarizes the performance results.

\section{Conclusion}
In this paper, we presented a QTBM formulation that combines a Bloch expansion of the OBCs and a low-rank approximation. We demonstrated that our approach allows us to compute the gate leakage transmission functions of realistic nanoelectronic devices with a 2D channel and gate lengths L$_g$ $>$ 25 nm. This achievement paves the way for investigating the influence of different dielectric layers on the leakage properties of ultra-scaled FETs.

\end{document}